\documentclass[%
 reprint,
groupedaddress,
 amsmath,amssymb,
 aps,
]{revtex4-1}

\usepackage{graphicx}%
\usepackage{xcolor}
\usepackage{dcolumn}%
\usepackage{bm}%

\newcommand{\avg}[1]{\ensuremath{\left<{#1}\right>}}

\newcommand{\bigo}[1]{\ensuremath{\mathcal{O}\left({#1}\right)}}

\newcommand{\kb}{\ensuremath{k_B}}
\newcommand{\gsf}{\ensuremath{\gamma_\text{SF}}}
\newcommand{\dd}{\ensuremath{\mathrm{d}}}

\begin{document}

\title{
Computing the Absolute Gibbs Free Energy in Atomistic Simulations:\\ Applications to Defects in Solids
}%

\author{Bingqing Cheng}
\email{bingqing.cheng@epfl.ch}
 \affiliation{Laboratory of Computational Science and Modeling, Institute of Materials, {\'E}cole Polytechnique F{\'e}d{\'e}rale de Lausanne, 1015 Lausanne, Switzerland}%

\author{Michele Ceriotti}
\affiliation{Laboratory of Computational Science and Modeling, Institute of Materials, {\'E}cole Polytechnique F{\'e}d{\'e}rale de Lausanne, 1015 Lausanne, Switzerland}%

\date{\today}%

\begin{abstract}
The Gibbs free energy is the fundamental thermodynamic potential underlying the relative stability of different states of matter under constant-pressure conditions.
However, computing this quantity from atomic-scale simulations is far from trivial. As a consequence, all too often the potential energy of the system is used as a proxy, 
 overlooking entropic and
anharmonic effects. 
Here we discuss a combination of different thermodynamic integration routes to obtain the absolute Gibbs free energy of a solid system starting from a harmonic reference state.
This approach enables the direct comparison between the free energy of different structures, circumventing the need to sample the transition paths between them.  
We showcase this thermodynamic integration scheme by computing the Gibbs free energy associated with a vacancy in BCC iron, and the intrinsic stacking fault free energy of nickel.
These examples highlight the pitfalls of estimating the free energy of crystallographic defects only using the minimum potential energy, which overestimates the vacancy free energy by 60\%{} and the stacking-fault energy by almost 300\%{} at temperatures close to the melting point. 
\end{abstract}

\maketitle

\section{\label{sec:0}Introduction}

Knowledge of the Gibbs free energy
is crucial in predicting the relative stability of different states of materials and molecules, and underlies a plethora of physical and chemical phenomena including
 phase diagrams, solubility, equilibrium concentration of defects, and so on.
However, free energy calculations in atomistic simulations are often technically challenging and/or computationally demanding. For this reason, in many cases -- particularly those involving solid phases -- the potential energy of the local minimum configuration associated with a given state is used as a simple proxy, and at times a harmonic correction is also included as the entropic term. 

One class of standard free energy techniques, such as metadynamics, umbrella sampling, and transition path sampling~\cite{laio2002escaping,torrie1977nonphysical,bolhuis2002transition},
relies on the concept that the phase space of a system can be divided into a number of states using a choice of reaction coordinates (or collective variables),
and the free energy difference between two states can then be computed by sampling both states, as well as the transition paths that connect them.

Another class of free energy methods concentrates on computing the ``absolute'' free energy of a system by performing a thermodynamic integration (TI)~\cite{tuckerman2010statistical,vega2008determination,ghiringhelli2005modeling}. 
The TI can be performed along a physical path for example along temperature or pressure, or via an unphysical path between the physical system and a reference system over a switching parameter $\lambda$.
For the latter route, the parameterized Hamiltonian can be taken to be $\mathcal{H}(\lambda)=(1-\lambda)\mathcal{H}_{ref}+\lambda \mathcal{H}$, where $\mathcal{H}$ is the actual Hamiltonian and $\mathcal{H}_{ref}$ is for the reference system with a known free energy. 
For example, the Helmoltz free energy of a crystal can be obtained by TI from an Einstein crystal, whose free energy can be expressed analytically, to the fully interacting system~\cite{polson1998calculation,anwar2003calculation,ghiringhelli2005modeling,ross+16prl,li2017computational}.
The Gibbs free energy of liquid water can be computed by following a thermodynamic path from a Lennard-Jones model to the real potential~\cite{sanz2004phase}.
However, each TI route is associated with different complications which have been
extensively discussed in the literature, such as the singularity around $T=0$ when integrating with respect to the system temperature~\cite{moustafa2015very}, and the pathological divergence of $d \mathcal{H}(\lambda)/ d \lambda$ that is often observed at the end points of the integral when switching between the real and the reference systems ~\cite{foiles1994evaluation,polson1998calculation,vega2008determination,schultz2016reformulation}.
As such, in general it is not trivial to find an optimal route for TI for a specific system.

The free energies associated with crystallographic defects (e.g. vacancies, dislocations, grain boundaries, surfaces, etc.) are extremely important in predicting the micro-structures and the properties of crystalline materials.
For example, the free energy of stacking faults is crucial in predicting the dislocation nucleation rate~\cite{warner2009origins},
the free energies associated with different surface reconstructions determine the surface phase diagram~\cite{valtiner2009temperature},
grain boundary free energy affects the rate of boundary migration~\cite{turnbull1951theory}.
However, computing the free energies associated with the defects is a particularly challenging problem, which reveals many of the shortcomings of standard free energy methods. 
Determining a physical or a virtual transition path to introduce or destroy
a defect inside a crystal is often complicated, ruling out techniques such as metadynamics or umbrella sampling. 
Thermodynamic integration over $\lambda$ using a harmonic reference often leads to divergences at high temperatures, when
diffusive and anharmonic behaviors become dominant~\cite{foiles1994evaluation}.
Due to these difficulties in free energy estimations, the defect free energies are usually approximated by just the potential energy of defects or by harmonic approximations~\cite{lesar1989finite,zimmerman2000generalized}.
Recently, a number of studies have revealed significant temperature dependency of the stacking fault free energy due to entropic effects~\cite{warner2009origins,ryu2011entropic,kim2014entropically}. 
It has also been shown that vacancy formation energy at high temperatures is strongly affected by anharmonicity~\cite{grabowski2009ab,glensk2014breakdown}.
Overall, the community is becoming more aware that the accuracy of the prediction from the minimum potential energies may deteriorate at high temperatures.

In this paper we combine a number of established thermodynamic integration methods (Figure~\ref{fig:tdi}) in order to devise a strategy for computing the absolute free energy of a solid system with and without crystallographic defects in an efficient, easy to implement, and generally-applicable way. 
We build a connection between the TI routes under the canonical (NVT) ensemble and the isothermal-isobaric (NPT) ensemble, in order to freely transform between the Helmoltz free energy and the Gibbs free energy of a system.
In addition, we incorporate several enhanced sampling methods and post-processing techniques to improve the overall statistical efficiency of the free energy estimation.
Finally, we showcase how to use thermodynamic integration to calculate the free energy of a vacancy in BCC iron and the intrinsic stacking fault free energy in FCC nickel.
We demonstrate that the anharmonic effects can significantly affect the free energies of defects in solids at high temperatures.

\section{\label{sec:1}Theory}

\begin{figure}
\includegraphics[width=0.45\textwidth]{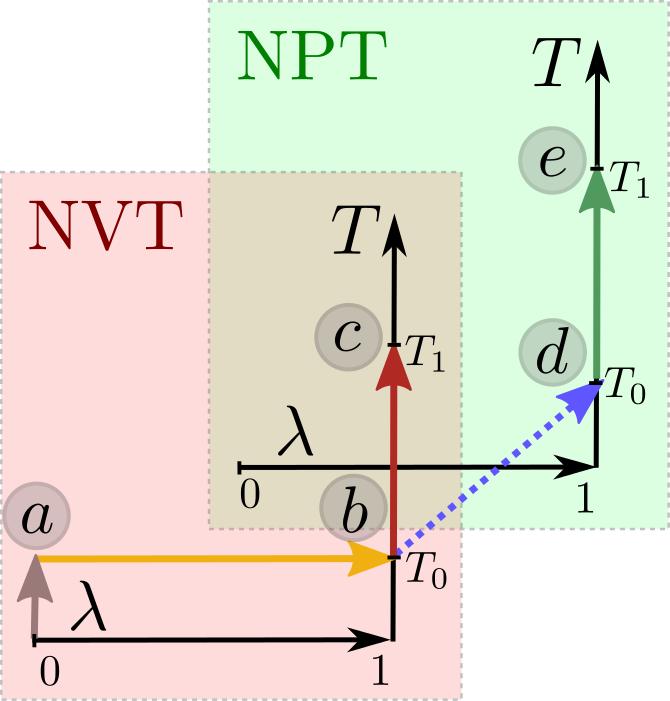}
\caption{
An illustration of the different thermodynamic integration routes employed in the present paper.
Under the canonical (NVT) ensemble,
the yellow arrow indicates the switching between an harmonic reference system ($\lambda=0$) and a real system ($\lambda=1$),
and the red arrow illustrates TI with respect to temperature.
The dashed blue arrow shows the  transformation  between  the Helmholtz free energy and the Gibbs free energy.
The green arrow denotes TI over temperature under the  isothermal-isobaric (NPT) ensemble.
}
\label{fig:tdi}
\end{figure}

The statistical-mecanical expression for the free energy of a system is closely related to the partition function, which in turn depends on the thermodynamic boundary conditions defining the ensemble.
Under the canonical (NVT) ensemble, the partition function of a bulk system that has $N$ \emph{indistinguishable} particles and is contained in a volume $V$ is given by~\cite{allen2012computer,tuckerman2010statistical}
\begin{equation}
  Q(N,V,T) = \dfrac{V^N}{\Lambda^{3N}N!}
  \int_{D(V)} \dd\textbf{q}
  \exp\left[-\dfrac{U(\textbf{q})}{\kb T}\right],
\end{equation}
where the potential energy $U$ is a function of the atomic coordinates $\textbf{q}=\{ \textbf{q}_{1\ldots N}\}$,
$D(V)$ denotes the spatial domain defined by the containing volume~\cite{tuckerman2010statistical},
and $\Lambda=\sqrt{2\pi\hbar^2/m \kb T}$ is the thermal de Broglie wavelength.
The expression for the Helmholtz free energy of the system is thus
\begin{multline}
  A(N,V,T) = 
  -\kb T \ln Q(N,V,T)  \\
  =- \kb T \ln \dfrac{V^N}{\Lambda^{3N}N!}
  -\kb T \ln  \int_{D(V)} \dd\textbf{q}
  \exp\left[-\dfrac{U(\textbf{q})}{\kb T}\right],
  \label{eq:A}
\end{multline}
where the first term in the last line is the free energy of an ideal gas.
When the isothermal-isobaric (NPT) ensemble is used instead,
the system can be characterized by the Gibbs free energy
\begin{equation}
    G(N,P,T) = -\kb T \ln \int dV \exp\left[-\dfrac{PV}{\kb T}\right]  \exp{\left[-\dfrac{A(N,V,T)}{\kb T}\right]}.
    \label{eq:gibbs}
\end{equation}

Computing directly the partition function for an arbitrary potential is impractical, and for this reason practical routes for computing $A$ or $G$ typically use a reference system for which the phase-space integral can be computed analytically, followed by one or more thermodynamic integration steps. 
Since TI does not require a smooth transformation of the atomic coordinates, but just the evaluation of free-energy derivatives as a function of a change in the thermodynamic conditions, many different paths can be used, and combined to obtain the most efficient protocol. 

In the case of a solid system, which is the main focus of the present work, we found it effective to take a harmonic crystal reference, and follow the TI routes illustrated in 
Figure~\ref{fig:tdi}.
In a nutshell,
for evaluating the Helmholtz free energy $A$ of a solid system,
we propose to first integrate along $\lambda$ 
between the harmonic and the real crystal and then do an
integration with respect to the temperature $T$,
which correspond to the yellow and the red arrows in Figure~\ref{fig:tdi}, respectively.
To calculate the Gibbs free energy $G$ of the system,
we first obtain the Helmholtz free energy at a low temperature,
switch from the NVT to the NPT ensemble,
and finally do TI along the temperature $T$
(the yellow, the blue, and the green arrows in Figure~\ref{fig:tdi}).
In the followings, we discuss in detail
how each step can be computed conveniently and efficiently.

Before we start the detailed discussion,
note that
in TI it is
advantageous and often necessary to constrain the center of mass (CM) of the system.
The Helmholtz free energy difference between the unconstrained and the constrained crystalline system 
under periodic boundary conditions can be
expressed as~\cite{polson2000finite}
\begin{equation}
  \Delta A_{cm}(N,V,T)
=-\kb T \Big(\ln \dfrac{V}{N}+\dfrac{3}{2}\ln N +\ln \dfrac{1}{\Lambda^3}\Big),
\label{eq:acm}
\end{equation}
which can be considered as a finite size effect.
Therefore, when we perform TI we focus solely on systems with fixed CM,
and at the end of the calculation the term $\Delta A_\text{CM}$ can be added to retrieve the free energy of the unconstrained system, although at times the influence may be negligible.
We will also discuss in more detail other  finite-size effects in Section ~\ref{sec:fse}.
In the other sections -- since we will always work under the constant-number-of-particles framework in TI --   we omit $N$ when denoting thermodynamic states.

\subsection{An absolute reference: the Helmoltz free energy of the Debye crystal}

Strictly speaking, only the relative free energy of a system with respect to a reference can be defined without any ambiguity. 
The ``absolute" free energy in this paper refers to the fact that the free energies of the chosen reference systems are analytic, and can be meaningfully compared between distinct reference systems including the ones that contain different numbers of particles.

A harmonically-coupled crystal of $N$ atoms with a constrained center of mass constitutes a convenient reference system for a solid (point $a$ in Figure~\ref{fig:tdi}). 
Taking the phonon frequency for the crystal to be
$\{ \omega_{i=1\ldots3N-3 } \}$,
\footnote{The zero-frequency translational modes are excluded due to the constraint on CM}
one can obtain an expression for the classical free energy of such a Debye crystal at the temperature $T_0$:
\begin{equation}
    A_\text{h} (V,T_0) =
\kb T_0
\sum_{i=1}^{3N-3}
\ln\dfrac{\hbar \omega_i}{\kb T_0}.
\label{eq:ahar}
\end{equation}
Note that from the standpoint of performing thermodynamic integration,
the reference can be any harmonic crystal that has the same number of particles as the real system.
For instance, one could even take a reference in which all particles are independently coupled to the lattice sites with a constant spring term (i.e. an Einstein crystal)~\cite{polson1998calculation,noya2008computing,ghiringhelli2005modeling}.
However,
for better statistical efficiency, it is better to choose a reference harmonic crystal that has the same frequency modes and equilibrium configuration as the real crystal,
both of which can be determined 
for example via local energy minimization followed by a diagonalization of the Hessian matrix~\cite{ayala1998identification}.

\subsection{The Helmoltz free energy of an anharmonic crystal}

Starting from a reference crystal ($a$) with a known free energy, one can obtain the Helmholtz free energy of the real crystal ($b$) using thermodynamic integration in the NVT ensemble, as indicated by the yellow arrow in Figure~\ref{fig:tdi}.
using a parameter $\lambda$ to perform the switch between the harmonic Hamiltonian $\mathcal{H}_\text{h}$ and the actual Hamiltonian $\mathcal{H}$.
In practice, one should run multiple simulations with the hamiltonian 
$\mathcal{H}(\lambda)=(1-\lambda)\mathcal{H}_{\text{h}}+\lambda \mathcal{H}$ at different values of $\lambda$, so as to switch between the harmonic Hamiltonian $\mathcal{H}_\text{h}$ and the actual Hamiltonian $\mathcal{H}$~\cite{polson1998calculation}.
The free energy of the real system with a fixed CM can then be evaluated using
\begin{equation}
  A(V,T_0)-A_{\text{h}}(V,T_0)=\int_0^1 d \lambda 
  \avg{U-U_{\text{h}}}_{V,T_0,\lambda},
  \label{eq:ti-l}
\end{equation}
where $\avg{\ldots}_{V,T_0,\lambda}$ denotes the ensemble average over NVT simulations using the Hamiltonian $H(\lambda)$.

In practice, to avoid severe statistical inefficiencies and singularities in the integral, one should perform this step at a low temperature $T_0$ when the system is quasi-harmonic and when diffusive or rotational degrees of freedom are completely frozen.
If $T_0$ is sufficiently low and
the real and the reference systems are very similar,
one also has the option to evaluate $A(V,T_0)-A_{\text{h}}(V,T_0)$ using the free energy perturbation method, eliminating the integration error altogether. 
One only needs to run simulations for the reference harmonic crystal,
and obtain the free energy of the real system using
\begin{equation}
    A(V, T_0) - A_{\text{h}}(V, T_0) =
    - k_B T_0 \ln\avg{
    \exp\left[-\dfrac{U-U_{\text{h}}}{k_B T_0}\right]}_{V,T_0,\lambda=0},
    \label{eq:gcut}
\end{equation}
where $\avg{\ldots}_{V,T_0,\lambda=0}$ denotes the ensemble average for the harmonic crystal at $T_0$ and $V$, and $U$ and $U_{\text{h}}$ denote the real and harmonic potentials, respectively. 
Note that in order to ensure the statistical efficiency of this perturbative approach, 
the standard deviation of $U - U_{\text{h}}$ has to be of the order of $\kb T_0$~\cite{ceriotti2011inefficiency}.

\subsection{\label{sec:ti-nvt}The Helmholtz free energy as a function of temperature}

Thermodynamic integration from a Debye crystal to the fully-anharmonic potential tends to become very inefficient as the temperature increases. For this reason, it is often useful to perform a thermodynamic integration with respect to temperature to from a low to a high temperature under the desired thermodynamic conditions. 
Let us start by discussing how to perform this step under the NVT ensemble, which is the process indicated by the red arrow ($b$ to $c$) in Figure~\ref{fig:tdi}.
The Helmholtz free energy of a system that has $N$ atoms and a fixed CM can be expressed by the well-known thermodynamic integration expression
\begin{equation}
    \dfrac{A{(V,T_1)}}{\kb T_1} = \dfrac{A{(V,T_0)}}{\kb T_0} 
    -\int_{T_0}^{T_1} \dfrac{\avg{U}_{V,T}+\avg{K}_{V,T}}{\kb T^2} dT,
    \label{eq:nvt-ti}
\end{equation}
where 
$\avg{U}_{V,T}$ and $\avg{K}_{V,T}$ are the ensemble averages of the potential energy and the kinetic energy, respectively.

One way to improve the convergence of Eqn.~\eqref{eq:nvt-ti}
is to consider that the ensemble average of the classical kinetic energy of a system that has $N$ atoms and a fixed CM is analytic: $\avg{K}_{V,T}=(3N-3)\kb T/2$. Furthermore, one can also consider that if the potential was harmonic, also $\avg{U}$ would take the same value. Thus, one can take
\begin{equation}
  \avg{\delta U}_{V,T} = \avg{U}_{V,T}-A(V,0)-(3N-3)\dfrac{\kb T}{2}
  \label{eq:deltau}
\end{equation}
that measures the temperature-dependent anharmonic part of the potential energy. Note that $A(V,0)=\avg{U}_{V,0}$.
After performing analytically some of the integrals, Eqn.~\eqref{eq:nvt-ti} becomes
\begin{multline}
    \dfrac{A(V,T_1)}{\kb T_1} = 
    \dfrac{A(V,0)}{\kb T_1}
    + 
    \dfrac{A(V,T_0) - A(V,0)}{\kb T_0} \\
    -(3N-3) \ln\dfrac{T_1}{T_0}
    - \int_{T_0}^{T_1} \dfrac{\avg{\delta U}_{V,T}}{\kb T^2} dT.
 \label{eq:a}
\end{multline}

In quasi-harmonic systems, one can further reduce the variance of the integrand. One can use again the analytical expression for  $\avg{K}_{V,T}$, together with the virial theorem, to write
\begin{equation}
 (3N-3)\frac{\kb T}{2} = \avg{K} = -\dfrac{1}{2} \sum_{i=1}^N \avg{\textbf{F}_i \textbf{q}_i}.
 \label{eq:virial}
\end{equation}
Here $\mathbf{q}_i$ and $\textbf{F}_i$ are the position of atom $i$ and the force vector acting on it.
Since the average force $\avg{\textbf{F}_i}$ is zero, one can also add an arbitrary reference position $\hat{\textbf{q}}_i$, and write~\cite{moustafa2015very,schultz2016reformulation}
\begin{equation}
  \avg{\delta U}_{V,T} = \avg{U+\dfrac{1}{2} \sum_{i=1}^N \textbf{F}_i (\textbf{q}_i-\hat{\textbf{q}}_i)}_{V,T}-A(V,0)
  \label{eq:du}
\end{equation}
If the potential energy surface is perfectly harmonic, and if one take $\hat{\textbf{q}}_i$ equal to the equilibrium position of atom $i$, it is easy to verify that the virial term would cancel completely the fluctuations in the potential energy.
Even if the potential is quasi-harmonic,
as long as it is not diffusive even at high temperatures, the use of the virial reference and Eqn.~\eqref{eq:du} can substantially improve the statistical efficiency in the estimation of $\avg{\delta U}_{V,T}$.
However, when the motion of atoms is strongly anharmonic or diffusive, an atom can start vibrating around a different equilibrium position, and in that case the statistical efficiency of the straightforward expression Eqn.~\eqref{eq:deltau} is better.

Besides improving the convergence of each temperature window, one can try to improve the accuracy and the efficiency of the TI procedure by choosing wisely the discretization points or, equivalently, by performing a change of variables that yields a smoother integrand~\cite{ceri-mark13jcp}. 
In this case, it is convenient to
perform a change of variables that ensures that the statistical error in the integrand is roughly
constant at all temperatures.
Assuming the temperature dependence of the fluctuations in the anharmonic potential energy is similar to its harmonic counterpart, i.e.
$\avg{\delta U^2}_{V,T}-\avg{\delta U}_{V,T}^2 \sim T$,
the required change of variable is 
$y=\ln(T/T_0)$, which transforms the integral into the form 
\begin{equation}
    \int_{T_0}^{T_1} \dfrac{\avg{\delta U}_{V,T}}{ T^2} dT
    =\int_{0}^{\ln(T_1/T_0)} 
     \frac{\avg{\delta U}_{V,T_0 e^y}}{T_0 e^y} dy.
    \label{eq:inty}
\end{equation}
In other words, one should select temperatures that are equally spaced in $\ln(T)$ in simulations.
Coincidentally, this selection is also optimal for performing replica exchanges between the systems at different temperatures~\cite{earl2005parallel} - which should be done whenever possible as it will greatly benefit statistical convergence.

Another advantage of performing parallel tempering is that it requires sufficient overlap between adjacent replicas at temperatures $T_{i}$ and $T_{i+1}$.
Under these circumstances,
$\avg{U}_{V,T}$ for $T_i<T<T_{i+1}$
can be evaluated via re-weighting, such as
\begin{equation}
\avg{U}_{V,T} = 
\dfrac{\avg{U\exp\Big[-\dfrac{U}{\kb}\big(\dfrac{1}{T}-\dfrac{1}{T_i}\big)\Big]}_{V,T_i}}
{\avg{
\exp\Big[-\dfrac{U}{\kb}\big(\dfrac{1}{T}-\dfrac{1}{T_i}\big)\Big]
}_{V,T_i}}.
\end{equation}
The integral in Eqn.~\eqref{eq:a} can thus be solved analytically to give an exact (within statistical uncertainty) expression for the contribution to the integral from the $[T_i,T_{i+1}]$ window:
\begin{multline}
  \dfrac{A{(V,T_{i+1})}}{\kb T_{i+1}} - \dfrac{A{(V,T_i)}}{\kb T_i} = \\
  - \dfrac{3N-3}{2} \ln\dfrac{T_{i+1}}{T_i}
  - \ln \avg{
  \exp\left[-\dfrac{U}{\kb}\big(\dfrac{1}{T_{i+1}}-\dfrac{1}{T_i}\big)\right]
  }_{V,T_i},
\end{multline}
which effectively turns the thermodynamic integration formalism into a sequence of free energy perturbations,
eliminating completely the integration error.

\subsection{From the Helmholtz free energy to the Gibbs free energy}

More often than not, the isothermal--isobaric ensemble (NPT) provides a more natural framework to describe the thermodynamic conditions of real systems than the NVT ensemble. 
However, the harmonic crystal ($a$ in Figure~\ref{fig:tdi}) that was used as the absolute constant-volume reference in previous sections does not extend naturally  to the NPT ensemble because its pressure is not well-defined (e.g. the Einstein crystal is a system of independent particles)~\cite{freitas2016nonequilibrium}.
As a result, it is not convenient to perform TI with respect to $\lambda$ from the reference crystal to the real crystal under the NPT ensemble.
One way to avoid NPT simulation involves performing multiple simulations at different constant volumes, and then compute the Gibbs free energy and the equilibrium volume of the system by evaluating explicitly the integral~\eqref{eq:gibbs}. This is often done using a harmonic expression for the free energy at the different volumes, leading to the so-called quasi-harmonic approximation (QHA)~\cite{xie1999first,ramirez2012quasi}.
Alternatively, the equilibrium volume can be computed by performing a single NPT simulation at the desired temperature, and then $A(\avg{V}_{P,T},T)$ is used as a proxy for $G(P,T)$~\cite{freitas2016nonequilibrium}.
In this section, we argue that the transformation between the Helmholtz free energy and the Gibbs free energy, which is the process marked by the dashed blue arrow in Figure~\ref{fig:tdi}, can be conducted rigorously. This process effectively allows us to convert at will between the two ensembles when performing thermodynamic integrations with respect to ensemble temperature.  

The expression for the Gibbs free energy of a system 
as an integral over the Helmholtz free energy is given by
Eqn.~\eqref{eq:gibbs}.
This expression can be combined with that for the distribution of volume fluctuations  for the system under the NPT ensemble
\begin{equation}
  \rho\left(V\middle|P,T\right) = 
  \dfrac{\exp\left[-\dfrac{PV}{\kb T}\right]  \exp{\left[-\dfrac{A(V,T)}{\kb T}\right]}}
  {\int dV \exp\left[-\dfrac{PV}{\kb T}\right]  \exp{\left[-\dfrac{A(V,T)}{\kb T}\right]}},
  \label{eq:rho},
\end{equation}
which is just the normalized probability of observing the system to have instantaneous volume $V$
in a simulation under constant P and T.
We can then write
\begin{equation}
  G(P,T) = A (V,T) + PV + \kb T \ln \rho\left(V\middle|P,T\right),
  \label{eq:atog}
\end{equation}
which is valid for arbitrary $V$.

In practice, one can run NPT simulations for a system and compute $\rho\left(V\middle|P,T\right)$ just by 
accumulating the histogram of the instantaneous volume of the system. 
After that,
one can select a volume $V$, preferably the one that maximizes $\rho\left(V\middle|P,T\right)$ for the sake of better statistical efficiency in the determination of $\rho\left(V\middle|P,T\right)$,
and compute $A(V,T)$ for the same system at that volume using the route $a$ to $b$ in Figure~\ref{fig:tdi}. 
Finally, the Gibbs free energy can be obtained applying Eqn.~\eqref{eq:atog}.

For a solid system,
in order to avoid residual strain and elastic energy,
one can vary the shape of the simulation cell instead of using
a fixed shape in NPT simulations under a hydrostatic pressure~\cite{parrinello1981polymorphic}.
To account for the degree of freedom associated with the variable cell, in this case Eqn.~\eqref{eq:atog} should be modified to read
\begin{equation}
  G(P,T) = A( \mathbf{h},T) + P \operatorname{det} (\mathbf{h}) + \kb T \ln \rho\left(\mathbf{h}\middle|P,T\right),
\end{equation}
where $\textbf{h}$ is a matrix that represents the dimensions of a simulation cell,
and $A( \mathbf{h},T)$ is the free energy of the system evaluated at constant cell dimensions.

\subsection{The Gibbs free energy as a function of temperature}

Having converted a harmonic-reference Helmoltz free-energy to a constant-pressure Gibbs free energy at a given temperature $T_0$, one can easily perform a thermodynamic integration over temperature in the NPT ensemble (a path indicated by the green arrow in Figure~\ref{fig:tdi}). 
For a system with $N$ atoms and a restricted CM, the expression reads
\begin{equation}
    \dfrac{G{(P,T_1)}}{\kb T_1} = \dfrac{G{(P,T_0)}}{\kb T_0} 
    -\int_{T_0}^{T_1} \dfrac{\avg{H}_{P,T}}{\kb T^2} dT,
    \label{eq:npt-ti}
\end{equation}
where $\avg{H}_{P,T}=\avg{U}_{P,T} + (3N-3)\dfrac{\kb T}{2} + P \avg{V}_{P,T}$ is the enthalpy.
Starting from this expression, one can apply all the techniques mentioned in Section~\ref{sec:ti-nvt},
for example one can take
\begin{equation}
  \avg{\delta H}_{P,T} = \avg{H}_{P,T}-G(P,0)-(3N-3)\dfrac{\kb T}{2},
\end{equation}
where $G(P,0)=\avg{U}_{P,0} + P \avg{V}_{P,0}$.
Doing the integration in Eqn.~\eqref{eq:npt-ti} explicitly leaves
\begin{multline}
    \dfrac{G(P,T_1)}{\kb T_1} = 
    \dfrac{G(P,0)}{\kb T_1}
    + 
    \dfrac{G(P,T_0) - G(P,0)}{\kb T_0} \\
    -(3N-3) \ln\dfrac{T_1}{T_0}
    - \int_{T_0}^{T_1} \dfrac{\avg{\delta H}_{P,T}}{\kb T^2} dT.
 \label{eq:g}
\end{multline}
In addition, one can also use the virial theorem (Eqn.~\eqref{eq:virial}), the change of variable in the integration (Eqn.~\eqref{eq:inty}),
and parallel tempering to further accelerate
the convergence.
When performing parallel tempering, one can eliminate the thermodynamic integration error by using a free-energy perturbation to compute the increment of $G$ between two replicas at temperatures $T_{i+1}$ and $T_i$:
\begin{multline}
  \dfrac{G{(P,T_{i+1})}}{\kb T_{i+1}} - \dfrac{G{(P,T_i)}}{\kb T_i} = \\ 
  - \dfrac{3N-3}{2} \ln\dfrac{T_{i+1}}{T_i}
  - \ln \avg{
  \exp\left[-\dfrac{U+PV}{\kb}\left(\dfrac{1}{T_{i+1}}-\dfrac{1}{T_{i}}\right)\right]
  }_{P,T_i}.
\end{multline}

\subsection{\label{sec:fse}Finite size effects}

Most of the time, one is interested in computing the free energy per atom of a bulk, infinite system, or the excess free energy of a defect in the dilute limit. 
An atomistic simulation, however, is inevitably restricted to a finite system size, which can result in deviations from the ideal case.
Many of these finite-system-size effects have been documented in the literature. 
First of all, in the limit of small system size, the free energy of the system is not an extensive quantity.
Taking the ideal gas part of the Helmholtz free energy $A_{id}(N,V,T)=- \kb T \ln \dfrac{V^N}{\Lambda^{3N}N!}$ in Eqn.~\eqref{eq:A} for example, one can see that
\begin{equation}
  \dfrac{A_{id}(N,V,T)}{N\kb T}
  =  1-\ln\dfrac{V}{N} + \ln\dfrac{1}{\Lambda^3}
  -\dfrac{\ln N}{2N} + \bigo{\dfrac{1}{N}}
\end{equation}
using Stirling's formula.
The leading $\ln N/N$ term is a well-documented finite size effect that reduces to zero in the thermodynamic limit~\cite{hoover1968entropy,polson2000finite}.
The Gibbs free energy per atom $G(N,P,T)/N$ as well as 
$\kb T \ln \rho\left(V\middle|P,T\right)/N$ in Eqn.~\eqref{eq:rho} also displays a similar dependence on $\ln N/N$.
Constraining the center of mass of the system in simulations also introduces a non-extensive correction to the free energy, 
\begin{equation}
  \Delta A_{cm}(N,V,T) = 
  A(N,V,T) - A_{cm}(N,V,T),
\end{equation}
where $A_{cm}$ denotes 
the Helmholtz free energy of the system with fixed center of mass~\cite{polson2000finite,vega2008determination}.
Fortunately, it is easy to correct for this part, because the expression for $\Delta A_{cm}$ in Eqn.~\eqref{eq:acm} is analytic and trivial to compute.
More subtle sources of finite size effects come from the cutoff of potentials, and from the discretization of the vibrational phonon 
spectrum due to the size of the supercell in simulations~\cite{freitas2016nonequilibrium}. 
To help with this issue, there are interpolation techniques that help accelerate the convergence of the computed phonon dispersion relation~\cite{giannozzi1991ab}.

It is worth stressing that for system sizes that can be reached easily in simulations using empirical force fields, finite-size effects may not be significant.
However, one should always be aware of their presence and check for system-size convergence, particularly in ab initio calculations where the number of atoms that can be simulated is highly restricted.
To minimize the impact of finite-size effects one should always compare free energies between systems of similar sizes, to benefit for a (partial) error cancellation.

\section{Application 1: vacancy free energy}
\subsection{Introduction}

A vacancy is a type of point defects in a crystal, in which an atom is removed from one of the lattice sites.
At any given temperature and pressure up to the solid-liquid coexistence line, an equilibrium concentration $\exp\left[-G_\text{v}/\kb T\right]$
of vacancies exists, where $G_\text{v}$ is the Gibbs free energy of a vacancy. Often, particularly in materials produced by fast quenching, a non-equilibrium concentration of vacancies can persist at low temperature, which can play an important role in technologically relevant solid-state transformations~\cite{poga+14prl}.

\subsection{Simulation details and results}

We studied a BCC iron system using a widely used EAM potential~\cite{mendelev2003development,bonny2009ternary}.
This potential was fitted with the BCC vacancy formation energy at 0K but lacks a thermally stable FCC phase~\cite{mendelev2003development,engin2008characterization}.
Iron exhibits phase transitions between BCC $\alpha$-iron, FCC $\gamma$ -iron and a BCC $\delta$-phase when increasing the temperature at ambient pressure, which are largely due to the magnetism of the material~\cite{engin2008characterization}.
Since the stabilization of the austenitic phase is due to quantum mechanical effects and this EAM potential does not reproduce it,
we neglect the FCC phase in the present study, and
performed the simulations considering a perfect BCC crystal (250 atoms) and a BCC crystal with a vacancy (249 atoms).
In all the simulations, the centers of mass of the systems were constrained.

At high temperatures, the computation of the free energy of the crystal with a vacancy is particularly problematic using the integration over $\lambda$ in Eqn.~\eqref{eq:ti-l}
due to the onset of diffusion in simulations~\cite{foiles1994evaluation}.
This difficulty is circumvented here as we performed the integration from the harmonic crystal to the real crystal at a low temperature $T_0=$100K and at the equilibrium cell size (the yellow arrow in Figure~\ref{fig:tdi}).
This harmonic crystal has the same phonon modes and Hessian matrix as the real system~\cite{ayala1998identification,ceriotti2014pi}.
Note that at this step the Helmholtz free energy of a crystal with a vacancy that sits at a fixed lattice site is computed, 
as the vacancy does not diffuse during the simulations at $T_0=$100K.
After that,
we switched to the NPT ensemble (the blue arrow in Figure~\ref{fig:tdi}),
and ran simulations at different temperatures and zero hydrostatic pressure, using stochastic velocity re-scaling for temperature control and the anisotropic Nose-Hoover barostat to vary the dimensions of the orthorhombic periodic supercell~\cite{bussi2007canonical,plim95jcp}. 
During this step, we obtained the temperature dependence of the free energies using Eqn.~\eqref{eq:npt-ti} (the green arrow in Figure~\ref{fig:tdi}).
At high temperatures, the vacancy does diffuse but diffusion does not change the values of $\avg{\delta H}_{P,T}$ compared with the case when the vacancy is fixed at one site, due to the translational symmetry of the lattice.
As such, the Gibbs free energy of a crystal with a fixed vacancy was obtained after integration using Eqn.~\eqref{eq:npt-ti}.

All the detailed procedures, the key data points,
annotated input files and python notebooks for data analysis are included in the supplemental material~\cite{SI}.
The absolute Gibbs free energies of the perfect BCC iron and the crystal with a fixed vacancy are plotted in Figure 2.

\begin{figure}
\includegraphics[width=0.5\textwidth]{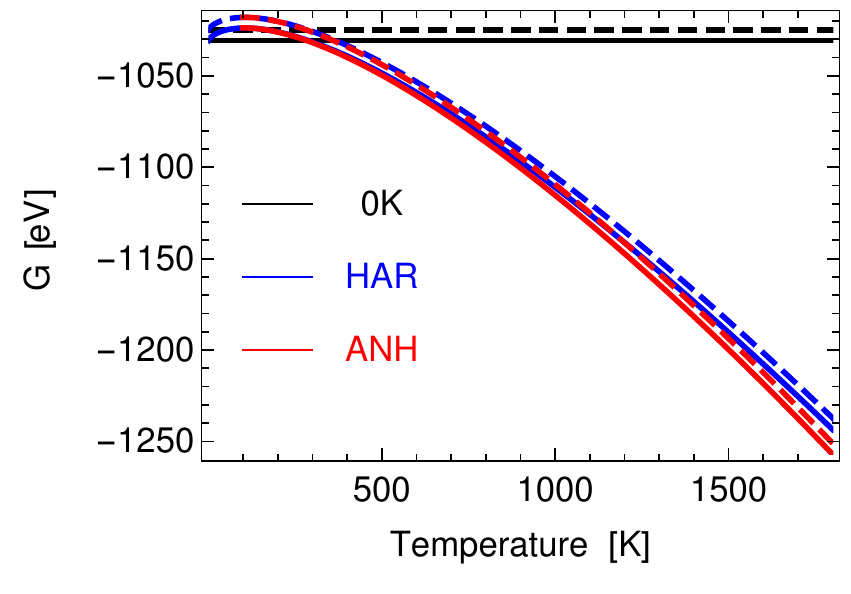}
\caption{
The solid and the dashed lines indicate the Gibbs free energy of the perfect BCC crystal and the crystal with a fixed vacancy, respectively.
The black lines are the estimations using just the potential energy at 0K,
the blues lines show the results from harmonic approximations,
and the red lines indicate the predictions that takes into account of all the anharmonic effects using the TI approach.
}
\label{fig:g-compare-v}
\end{figure}

\subsection{Gibbs free energy of a vacancy}

\begin{figure}
\includegraphics[width=0.5\textwidth]{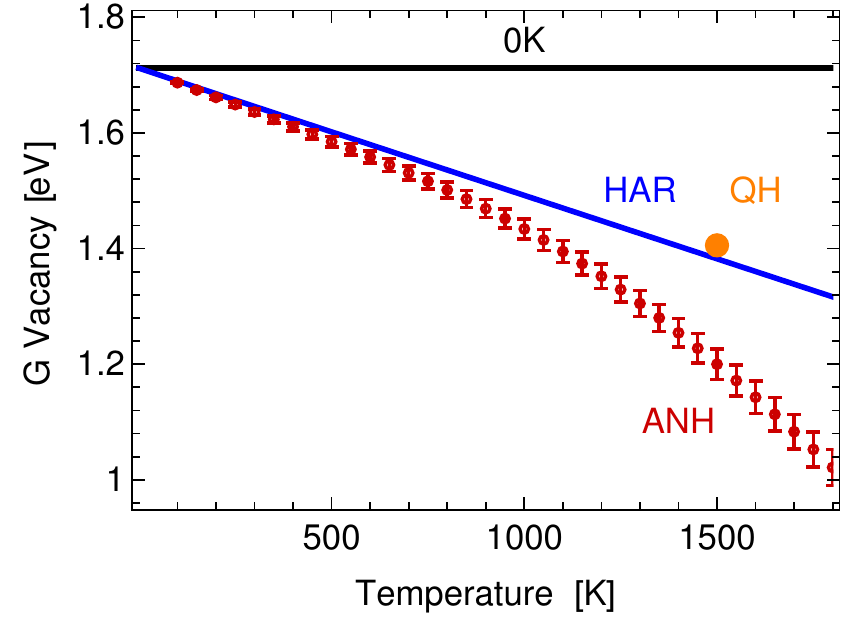}
\caption{
The Gibbs free energy associated with a fixed vacancy in BCC iron estimated using potential energy difference (PE) at 0K, the harmonic approximation (HAR), and the thermodynamic integration method that considers anharmonicity (ANH).
QH indicates the quasi-harmonic approximation using the equilibrium configuration at 1500K.
Statistical uncertainties are indicated by the error bars.
}
\label{fig:ve}
\end{figure}

The Gibbs free energy of a fixed vacancy in the crystal can be expressed as
\begin{equation}
  G_\text{v} = G_\text{{vacancy}} - \dfrac{N_\text{{vacancy}}}{N_{\text{perfect}}} G_{\text{perfect}}.
\end{equation}
We used three different methods to estimate this quantity -- namely a minimum-potential energy calculation, a harmonic free-energy estimate and the fully anharmonic TI -- and plot the results in Figure~\ref{fig:ve} as a function of temperature.
Notice that the defect free energy is tiny compared with the magnitude of the absolute free energy of both systems.
The difference between the predictions from the harmonic approximation and the TI is largely negligible at low temperatures,
but becomes significant when the temperature approaches the melting point 1772K for this EAM potential system~\cite{mendelev2003development}.
To investigate whether this difference stems from a shift in the phonon spectra due to lattice expansion, or from anharmocity,
we analyzed the vibrational modes  $\{ \omega'_{i=1\ldots3N-3 } \}$ using the equilibrium configuration of each system at 1500K,
computed the vacancy free energy under this quasi-harmonic approximation,
and plotted the result as the orange dot in Figure~\ref{fig:ve}.
It can be seen that the harmonic contribution at 1500K cannot explain the difference,
and therefore the difference is mainly due to anharmonic effects.

\section{Application 2: stacking fault free energy}
\subsection{Introduction}

A stacking fault (SF) is a defect in the planar stacking sequence of atoms in a crystal. While the perfect stacking for FCC crystals metals on the $\{111\}$ plane is $ABC | ABC | ABC |$, an intrinsic stacking fault changes the arrangement to $ABC | AB | ABC |$ as if one plane had been removed. 
The stacking fault free energy ($\gsf$) measures the free energy increase that is associated with the presence of the intrinsic stacking fault plane. 
The stacking fault free energy has a significant effect on the plastic deformation behavior of crystalline materials. For example,  metals with a low $\gsf$ form more stacking faults and twins, and more extended partial dislocations which have reduced mobility~\cite{allain2004correlations,van2004stacking,yamakov2004deformation}. 
Stacking faults also provide a strong barrier to dislocation gliding~\cite{allain2004correlations,van2004stacking,yamakov2004deformation}.

\subsection{Simulation details and results}

A widely used EAM potential that describes the Ni-Ni interactions was employed~\cite{voter1986accurate,bonny2009ternary}.
Note that different EAM potentials usually yield quite divergent estimates for the stacking fault energy,
and a comprehensive comparison between them together with a few DFT results can be found in Ref.~\cite{zimmerman2000generalized}.
As such, the purpose of this application is to highlight the temperature dependence of $\gsf$ for a generic metallic system
rather than a quantitative analysis for pure Ni, which would require a more reliable interatomic potential.

\begin{figure}
\includegraphics[width=0.5\textwidth]{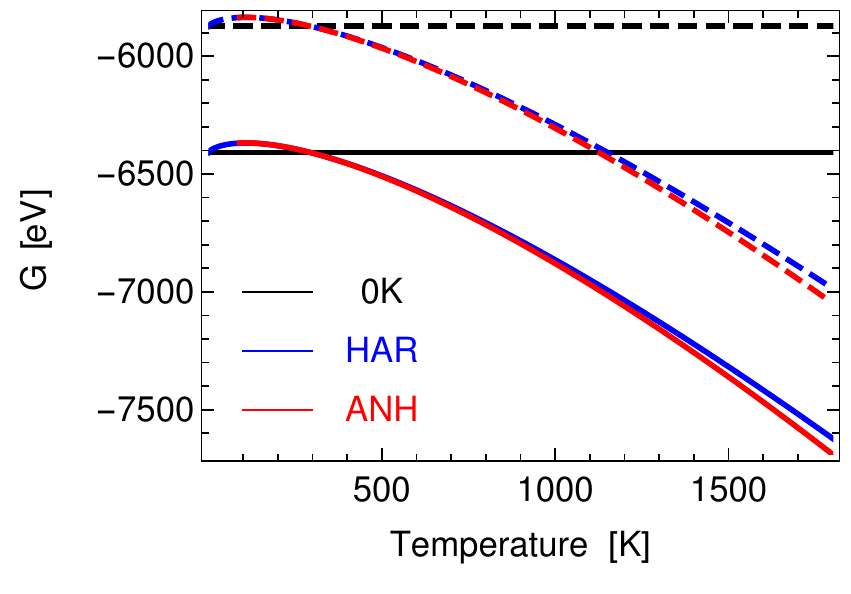}
\caption{
The solid and the dashed lines indicate the absolute Gibbs free energy of the perfect FCC crystal and the crystal with a stacking fault, respectively.
The black lines are the estimations using just the potential energy at 0K,
the blues lines shows the results from harmonic approximations,
and the red lines indicate the predictions that takes into account of all the anharmonic effects using thermodynamic integration.
The statistical errors are minute at the scale of the plot.
}
\label{fig:g-compare}
\end{figure}

We performed the simulations of a perfect FCC crystal and of a FCC crystal with a stacking fault layer separately. The perfect FCC crystal used in simulations has 12 layers of $\{111\}$ planes, and a total of $N_{\text{perfect}}=1440$ atoms. The defective crystal has 11 layers with the stacking $ABC | ABC | AB | ABC |$,
and a total of $N_\text{{sf}}=1320$ atoms.
We calculated separately the Gibbs free energy for both systems, and the results are reported in Figure~\ref{fig:g-compare}.
The centers of the mass of the systems were constrained in all the simulations.
All the detailed procedures and the key data points used in the calculations are included in the supplemental material~\cite{SI},
along with annotated input files and python notebooks for data analysis.
In brief,
we first computed the Helmoltz free energy of a real system at $T_0=$90K by thermodynamic integration with respect to $\lambda$ starting from a reference system using Eqn.~\eqref{eq:ti-l}.
The reference system is a harmonic crystal that has the same phonon modes and the Hessian matrix as the real system at the minimum potential energy~\cite{ayala1998identification,ceriotti2014pi},
and the simulation cell is kept constant at the equilibrium size of the real system at 90K.
Afterwards, 
at temperatures that are equally spaced in $\ln(T)$, independent molecular dynamics simulations were performed under the NPT ensemble with zero external stress~\cite{bussi2007canonical,plim95jcp},
in order to compute
the anharmonic contributions $\avg{\delta H}_{P,T}$ and
employ thermodynamic integration with respect to $T$ using Eqn.~\eqref{eq:g}. 
We also found the use of the virial theorem to significantly reduce the fluctuations in the estimation of $\avg{\delta H}_{P,T}$ when we select $\hat{\textbf{q}}$ to be the equilibrium position of atoms in Eqn.~\eqref{eq:du}.

\subsection{Stacking fault free energy estimations}

The free energy excess ($\gsf \times \text{Area}$) that is associated with the stacking fault plane with a surface area equal to the cross section of the simulation supercell is just the difference between the free energies of a crystal with a stacking fault and a perfect bulk FCC crystal that have the same number of atoms: 
\begin{equation}
  \gsf \times Area = G_\text{{sf}} - \dfrac{N_\text{{sf}}}{N_\text{{perfect}}} G_\text{{perfect}}.
\end{equation}

\begin{figure}
\includegraphics[width=0.5\textwidth]{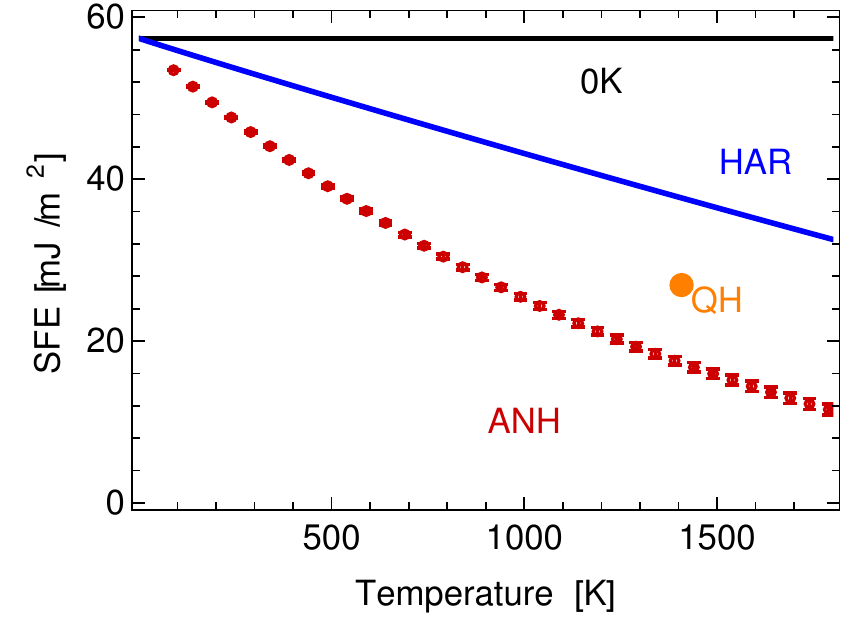}
\caption{
The area-specific stacking fault free energy estimated from potential energy difference at 0K, the harmonic approximation (HAR), and the thermodynamic integration method that considers anharmonicity (ANH).
QH indicates the quasi-harmonic approximation using the equilibrium configuration at 1408K.
Statistical uncertainties are indicated by the error bars.
}
\label{fig:sfe}
\end{figure}

We used three different methods to estimate $\gsf(T)$, and juxtaposed them in Figure~\ref{fig:sfe}.
We found the estimations from just the potential energy at 0K and the harmonic approximation are only accurate at very low temperatures.
The prediction of the quasi-harmonic approximation using the equilibrium configuration at 1408K is more accurate than the harmonic approximation,
but still falls short.
The inclusion of all the entropic and anharmonic effects dramatically decreases the stacking fault free energy at high temperatures,
making the magnitude of the stacking fault free energy at $T>$1200K less than one third of the value predicted from the 0K potential energies.

\section{Conclusions}

In the present study we have discussed a number of thermodynamic integration routes,
and devised a strategy to combine them in a way that makes the best use of the similarity between a crystalline system and a harmonic Debye crystal at low temperatures,
and at the same time fully take into account the anharmonic effects.
Furthermore, we discuss how to convert between free energy values estimated under constant-volume and constant-pressure thermodynamic ensembles, and
incorporate several techniques to enhance the efficiency and accuracy of free energy estimation.

The thermodynamic integration method does not rely on any approximation for anharmonic effects,
thus it can also serve as a benchmark for other approximation methods such as
quasi-harmonic approximations,
self consistent phonons~\cite{hooton1955li,brow+13jcp,errea2014anharmonic},
the Green’s function approach~\cite{campana2006practical}.
In addition, it is also possible to include the oft-neglected quantum nuclear effects by adding an extra TI route from classical to quantum mechanical system by employing path-integral molecular dynamics~\cite{ross+15jpcl,ross+16prl}.

We computed the free energy of a vacancy in BCC iron and the stacking fault free energy of FCC nickel.
In both cases, the anharmonic effects significantly lower the free energy of defects close to the melting point.
These results reinforce the notion that it is not sufficient to use potential energy minima, or the associated harmonic free energy, to estimate the stability of crystallographic defects at high temperatures.
In the supplementary materials~\cite{SI}, We have included all the input files, work-flows, and data analysis routines that were employed to produce the results discussed in the present paper, hoping to encourage and facilitate future efforts that aim to use thermodynamic integration methods to accurately predict the defect concentrations, phase diagrams, or any of the many materials properties that are determined by the Gibbs free energy.

\begin{acknowledgements}
We thank William Curtin for insightful comments on an early version of the manuscript. 
We also thank David Wilkins and Benjamin Helfrecht for critically reading the manuscript.
BC acknowledges funding from the Swiss National
Science Foundation (project ID 200021-159896). MC acknowledges financial support from  the CCMX Project ``AM$^3$''.
\end{acknowledgements}


\begin{thebibliography}{54}%
\makeatletter
\providecommand \@ifxundefined [1]{%
 \@ifx{#1\undefined}
}%
\providecommand \@ifnum [1]{%
 \ifnum #1\expandafter \@firstoftwo
 \else \expandafter \@secondoftwo
 \fi
}%
\providecommand \@ifx [1]{%
 \ifx #1\expandafter \@firstoftwo
 \else \expandafter \@secondoftwo
 \fi
}%
\providecommand \natexlab [1]{#1}%
\providecommand \enquote  [1]{``#1''}%
\providecommand \bibnamefont  [1]{#1}%
\providecommand \bibfnamefont [1]{#1}%
\providecommand \citenamefont [1]{#1}%
\providecommand \href@noop [0]{\@secondoftwo}%
\providecommand \href [0]{\begingroup \@sanitize@url \@href}%
\providecommand \@href[1]{\@@startlink{#1}\@@href}%
\providecommand \@@href[1]{\endgroup#1\@@endlink}%
\providecommand \@sanitize@url [0]{\catcode `\\12\catcode `\$12\catcode
  `\&12\catcode `\#12\catcode `\^12\catcode `\_12\catcode `\%12\relax}%
\providecommand \@@startlink[1]{}%
\providecommand \@@endlink[0]{}%
\providecommand \url  [0]{\begingroup\@sanitize@url \@url }%
\providecommand \@url [1]{\endgroup\@href {#1}{\urlprefix }}%
\providecommand \urlprefix  [0]{URL }%
\providecommand \Eprint [0]{\href }%
\providecommand \doibase [0]{http://dx.doi.org/}%
\providecommand \selectlanguage [0]{\@gobble}%
\providecommand \bibinfo  [0]{\@secondoftwo}%
\providecommand \bibfield  [0]{\@secondoftwo}%
\providecommand \translation [1]{[#1]}%
\providecommand \BibitemOpen [0]{}%
\providecommand \bibitemStop [0]{}%
\providecommand \bibitemNoStop [0]{.\EOS\space}%
\providecommand \EOS [0]{\spacefactor3000\relax}%
\providecommand \BibitemShut  [1]{\csname bibitem#1\endcsname}%
\let\auto@bib@innerbib\@empty
\bibitem [{\citenamefont {Laio}\ and\ \citenamefont
  {Parrinello}(2002)}]{laio2002escaping}%
  \BibitemOpen
  \bibfield  {author} {\bibinfo {author} {\bibfnamefont {A.}~\bibnamefont
  {Laio}}\ and\ \bibinfo {author} {\bibfnamefont {M.}~\bibnamefont
  {Parrinello}},\ }\href@noop {} {\bibfield  {journal} {\bibinfo  {journal}
  {Proceedings of the National Academy of Sciences}\ }\textbf {\bibinfo
  {volume} {99}},\ \bibinfo {pages} {12562} (\bibinfo {year}
  {2002})}\BibitemShut {NoStop}%
\bibitem [{\citenamefont {Torrie}\ and\ \citenamefont
  {Valleau}(1977)}]{torrie1977nonphysical}%
  \BibitemOpen
  \bibfield  {author} {\bibinfo {author} {\bibfnamefont {G.~M.}\ \bibnamefont
  {Torrie}}\ and\ \bibinfo {author} {\bibfnamefont {J.~P.}\ \bibnamefont
  {Valleau}},\ }\href@noop {} {\bibfield  {journal} {\bibinfo  {journal}
  {Journal of Computational Physics}\ }\textbf {\bibinfo {volume} {23}},\
  \bibinfo {pages} {187} (\bibinfo {year} {1977})}\BibitemShut {NoStop}%
\bibitem [{\citenamefont {Bolhuis}\ \emph {et~al.}(2002)\citenamefont
  {Bolhuis}, \citenamefont {Chandler}, \citenamefont {Dellago},\ and\
  \citenamefont {Geissler}}]{bolhuis2002transition}%
  \BibitemOpen
  \bibfield  {author} {\bibinfo {author} {\bibfnamefont {P.~G.}\ \bibnamefont
  {Bolhuis}}, \bibinfo {author} {\bibfnamefont {D.}~\bibnamefont {Chandler}},
  \bibinfo {author} {\bibfnamefont {C.}~\bibnamefont {Dellago}}, \ and\
  \bibinfo {author} {\bibfnamefont {P.~L.}\ \bibnamefont {Geissler}},\
  }\href@noop {} {\bibfield  {journal} {\bibinfo  {journal} {Annual review of
  physical chemistry}\ }\textbf {\bibinfo {volume} {53}},\ \bibinfo {pages}
  {291} (\bibinfo {year} {2002})}\BibitemShut {NoStop}%
\bibitem [{\citenamefont {Tuckerman}(2010)}]{tuckerman2010statistical}%
  \BibitemOpen
  \bibfield  {author} {\bibinfo {author} {\bibfnamefont {M.}~\bibnamefont
  {Tuckerman}},\ }\href@noop {} {\emph {\bibinfo {title} {Statistical
  mechanics: theory and molecular simulation}}}\ (\bibinfo  {publisher} {Oxford
  University Press},\ \bibinfo {year} {2010})\BibitemShut {NoStop}%
\bibitem [{\citenamefont {Vega}\ \emph {et~al.}(2008)\citenamefont {Vega},
  \citenamefont {Sanz}, \citenamefont {Abascal},\ and\ \citenamefont
  {Noya}}]{vega2008determination}%
  \BibitemOpen
  \bibfield  {author} {\bibinfo {author} {\bibfnamefont {C.}~\bibnamefont
  {Vega}}, \bibinfo {author} {\bibfnamefont {E.}~\bibnamefont {Sanz}}, \bibinfo
  {author} {\bibfnamefont {J.}~\bibnamefont {Abascal}}, \ and\ \bibinfo
  {author} {\bibfnamefont {E.}~\bibnamefont {Noya}},\ }\href@noop {} {\bibfield
   {journal} {\bibinfo  {journal} {Journal of Physics: Condensed Matter}\
  }\textbf {\bibinfo {volume} {20}},\ \bibinfo {pages} {153101} (\bibinfo
  {year} {2008})}\BibitemShut {NoStop}%
\bibitem [{\citenamefont {Ghiringhelli}\ \emph {et~al.}(2005)\citenamefont
  {Ghiringhelli}, \citenamefont {Los}, \citenamefont {Meijer}, \citenamefont
  {Fasolino},\ and\ \citenamefont {Frenkel}}]{ghiringhelli2005modeling}%
  \BibitemOpen
  \bibfield  {author} {\bibinfo {author} {\bibfnamefont {L.~M.}\ \bibnamefont
  {Ghiringhelli}}, \bibinfo {author} {\bibfnamefont {J.~H.}\ \bibnamefont
  {Los}}, \bibinfo {author} {\bibfnamefont {E.~J.}\ \bibnamefont {Meijer}},
  \bibinfo {author} {\bibfnamefont {A.}~\bibnamefont {Fasolino}}, \ and\
  \bibinfo {author} {\bibfnamefont {D.}~\bibnamefont {Frenkel}},\ }\href@noop
  {} {\bibfield  {journal} {\bibinfo  {journal} {Physical review letters}\
  }\textbf {\bibinfo {volume} {94}},\ \bibinfo {pages} {145701} (\bibinfo
  {year} {2005})}\BibitemShut {NoStop}%
\bibitem [{\citenamefont {Polson}\ and\ \citenamefont
  {Frenkel}(1998)}]{polson1998calculation}%
  \BibitemOpen
  \bibfield  {author} {\bibinfo {author} {\bibfnamefont {J.~M.}\ \bibnamefont
  {Polson}}\ and\ \bibinfo {author} {\bibfnamefont {D.}~\bibnamefont
  {Frenkel}},\ }\href@noop {} {\bibfield  {journal} {\bibinfo  {journal} {The
  Journal of chemical physics}\ }\textbf {\bibinfo {volume} {109}},\ \bibinfo
  {pages} {318} (\bibinfo {year} {1998})}\BibitemShut {NoStop}%
\bibitem [{\citenamefont {Anwar}\ \emph {et~al.}(2003)\citenamefont {Anwar},
  \citenamefont {Frenkel},\ and\ \citenamefont {Noro}}]{anwar2003calculation}%
  \BibitemOpen
  \bibfield  {author} {\bibinfo {author} {\bibfnamefont {J.}~\bibnamefont
  {Anwar}}, \bibinfo {author} {\bibfnamefont {D.}~\bibnamefont {Frenkel}}, \
  and\ \bibinfo {author} {\bibfnamefont {M.~G.}\ \bibnamefont {Noro}},\
  }\href@noop {} {\bibfield  {journal} {\bibinfo  {journal} {The Journal of
  chemical physics}\ }\textbf {\bibinfo {volume} {118}},\ \bibinfo {pages}
  {728} (\bibinfo {year} {2003})}\BibitemShut {NoStop}%
\bibitem [{\citenamefont {Rossi}\ \emph {et~al.}(2016)\citenamefont {Rossi},
  \citenamefont {Gasparotto},\ and\ \citenamefont {Ceriotti}}]{ross+16prl}%
  \BibitemOpen
  \bibfield  {author} {\bibinfo {author} {\bibfnamefont {M.}~\bibnamefont
  {Rossi}}, \bibinfo {author} {\bibfnamefont {P.}~\bibnamefont {Gasparotto}}, \
  and\ \bibinfo {author} {\bibfnamefont {M.}~\bibnamefont {Ceriotti}},\
  }\href@noop {} {\bibfield  {journal} {\bibinfo  {journal} {Phys. Rev. Lett.}\
  }\textbf {\bibinfo {volume} {117}},\ \bibinfo {pages} {115702} (\bibinfo
  {year} {2016})}\BibitemShut {NoStop}%
\bibitem [{\citenamefont {Li}\ \emph {et~al.}(2017)\citenamefont {Li},
  \citenamefont {Totton},\ and\ \citenamefont {Frenkel}}]{li2017computational}%
  \BibitemOpen
  \bibfield  {author} {\bibinfo {author} {\bibfnamefont {L.}~\bibnamefont
  {Li}}, \bibinfo {author} {\bibfnamefont {T.}~\bibnamefont {Totton}}, \ and\
  \bibinfo {author} {\bibfnamefont {D.}~\bibnamefont {Frenkel}},\ }\href@noop
  {} {\bibfield  {journal} {\bibinfo  {journal} {The Journal of Chemical
  Physics}\ }\textbf {\bibinfo {volume} {146}},\ \bibinfo {pages} {214110}
  (\bibinfo {year} {2017})}\BibitemShut {NoStop}%
\bibitem [{\citenamefont {Sanz}\ \emph {et~al.}(2004)\citenamefont {Sanz},
  \citenamefont {Vega}, \citenamefont {Abascal},\ and\ \citenamefont
  {MacDowell}}]{sanz2004phase}%
  \BibitemOpen
  \bibfield  {author} {\bibinfo {author} {\bibfnamefont {E.}~\bibnamefont
  {Sanz}}, \bibinfo {author} {\bibfnamefont {C.}~\bibnamefont {Vega}}, \bibinfo
  {author} {\bibfnamefont {J.}~\bibnamefont {Abascal}}, \ and\ \bibinfo
  {author} {\bibfnamefont {L.}~\bibnamefont {MacDowell}},\ }\href@noop {}
  {\bibfield  {journal} {\bibinfo  {journal} {Physical review letters}\
  }\textbf {\bibinfo {volume} {92}},\ \bibinfo {pages} {255701} (\bibinfo
  {year} {2004})}\BibitemShut {NoStop}%
\bibitem [{\citenamefont {Moustafa}\ \emph {et~al.}(2015)\citenamefont
  {Moustafa}, \citenamefont {Schultz},\ and\ \citenamefont
  {Kofke}}]{moustafa2015very}%
  \BibitemOpen
  \bibfield  {author} {\bibinfo {author} {\bibfnamefont {S.~G.}\ \bibnamefont
  {Moustafa}}, \bibinfo {author} {\bibfnamefont {A.~J.}\ \bibnamefont
  {Schultz}}, \ and\ \bibinfo {author} {\bibfnamefont {D.~A.}\ \bibnamefont
  {Kofke}},\ }\href@noop {} {\bibfield  {journal} {\bibinfo  {journal}
  {Physical Review E}\ }\textbf {\bibinfo {volume} {92}},\ \bibinfo {pages}
  {043303} (\bibinfo {year} {2015})}\BibitemShut {NoStop}%
\bibitem [{\citenamefont {Foiles}(1994)}]{foiles1994evaluation}%
  \BibitemOpen
  \bibfield  {author} {\bibinfo {author} {\bibfnamefont {S.~M.}\ \bibnamefont
  {Foiles}},\ }\href@noop {} {\bibfield  {journal} {\bibinfo  {journal}
  {Physical Review B}\ }\textbf {\bibinfo {volume} {49}},\ \bibinfo {pages}
  {14930} (\bibinfo {year} {1994})}\BibitemShut {NoStop}%
\bibitem [{\citenamefont {Schultz}\ \emph {et~al.}(2016)\citenamefont
  {Schultz}, \citenamefont {Moustafa}, \citenamefont {Lin}, \citenamefont
  {Weinstein},\ and\ \citenamefont {Kofke}}]{schultz2016reformulation}%
  \BibitemOpen
  \bibfield  {author} {\bibinfo {author} {\bibfnamefont {A.~J.}\ \bibnamefont
  {Schultz}}, \bibinfo {author} {\bibfnamefont {S.~G.}\ \bibnamefont
  {Moustafa}}, \bibinfo {author} {\bibfnamefont {W.}~\bibnamefont {Lin}},
  \bibinfo {author} {\bibfnamefont {S.~J.}\ \bibnamefont {Weinstein}}, \ and\
  \bibinfo {author} {\bibfnamefont {D.~A.}\ \bibnamefont {Kofke}},\ }\href@noop
  {} {\bibfield  {journal} {\bibinfo  {journal} {Journal of chemical theory and
  computation}\ }\textbf {\bibinfo {volume} {12}},\ \bibinfo {pages} {1491}
  (\bibinfo {year} {2016})}\BibitemShut {NoStop}%
\bibitem [{\citenamefont {Warner}\ and\ \citenamefont
  {Curtin}(2009)}]{warner2009origins}%
  \BibitemOpen
  \bibfield  {author} {\bibinfo {author} {\bibfnamefont {D.~H.}\ \bibnamefont
  {Warner}}\ and\ \bibinfo {author} {\bibfnamefont {W.}~\bibnamefont
  {Curtin}},\ }\href@noop {} {\bibfield  {journal} {\bibinfo  {journal} {Acta
  Materialia}\ }\textbf {\bibinfo {volume} {57}},\ \bibinfo {pages} {4267}
  (\bibinfo {year} {2009})}\BibitemShut {NoStop}%
\bibitem [{\citenamefont {Valtiner}\ \emph {et~al.}(2009)\citenamefont
  {Valtiner}, \citenamefont {Todorova}, \citenamefont {Grundmeier},\ and\
  \citenamefont {Neugebauer}}]{valtiner2009temperature}%
  \BibitemOpen
  \bibfield  {author} {\bibinfo {author} {\bibfnamefont {M.}~\bibnamefont
  {Valtiner}}, \bibinfo {author} {\bibfnamefont {M.}~\bibnamefont {Todorova}},
  \bibinfo {author} {\bibfnamefont {G.}~\bibnamefont {Grundmeier}}, \ and\
  \bibinfo {author} {\bibfnamefont {J.}~\bibnamefont {Neugebauer}},\
  }\href@noop {} {\bibfield  {journal} {\bibinfo  {journal} {Physical review
  letters}\ }\textbf {\bibinfo {volume} {103}},\ \bibinfo {pages} {065502}
  (\bibinfo {year} {2009})}\BibitemShut {NoStop}%
\bibitem [{\citenamefont {Turnbull}(1951)}]{turnbull1951theory}%
  \BibitemOpen
  \bibfield  {author} {\bibinfo {author} {\bibfnamefont {D.}~\bibnamefont
  {Turnbull}},\ }\href@noop {} {\bibfield  {journal} {\bibinfo  {journal}
  {Trans. Aime}\ }\textbf {\bibinfo {volume} {191}},\ \bibinfo {pages} {661}
  (\bibinfo {year} {1951})}\BibitemShut {NoStop}%
\bibitem [{\citenamefont {LeSar}\ \emph {et~al.}(1989)\citenamefont {LeSar},
  \citenamefont {Najafabadi},\ and\ \citenamefont
  {Srolovitz}}]{lesar1989finite}%
  \BibitemOpen
  \bibfield  {author} {\bibinfo {author} {\bibfnamefont {R.}~\bibnamefont
  {LeSar}}, \bibinfo {author} {\bibfnamefont {R.}~\bibnamefont {Najafabadi}}, \
  and\ \bibinfo {author} {\bibfnamefont {D.}~\bibnamefont {Srolovitz}},\
  }\href@noop {} {\bibfield  {journal} {\bibinfo  {journal} {Physical Review
  Letters}\ }\textbf {\bibinfo {volume} {63}},\ \bibinfo {pages} {624}
  (\bibinfo {year} {1989})}\BibitemShut {NoStop}%
\bibitem [{\citenamefont {Zimmerman}\ \emph {et~al.}(2000)\citenamefont
  {Zimmerman}, \citenamefont {Gao},\ and\ \citenamefont
  {Abraham}}]{zimmerman2000generalized}%
  \BibitemOpen
  \bibfield  {author} {\bibinfo {author} {\bibfnamefont {J.~A.}\ \bibnamefont
  {Zimmerman}}, \bibinfo {author} {\bibfnamefont {H.}~\bibnamefont {Gao}}, \
  and\ \bibinfo {author} {\bibfnamefont {F.~F.}\ \bibnamefont {Abraham}},\
  }\href@noop {} {\bibfield  {journal} {\bibinfo  {journal} {Modelling and
  Simulation in Materials Science and Engineering}\ }\textbf {\bibinfo {volume}
  {8}},\ \bibinfo {pages} {103} (\bibinfo {year} {2000})}\BibitemShut {NoStop}%
\bibitem [{\citenamefont {Ryu}\ \emph {et~al.}(2011)\citenamefont {Ryu},
  \citenamefont {Kang},\ and\ \citenamefont {Cai}}]{ryu2011entropic}%
  \BibitemOpen
  \bibfield  {author} {\bibinfo {author} {\bibfnamefont {S.}~\bibnamefont
  {Ryu}}, \bibinfo {author} {\bibfnamefont {K.}~\bibnamefont {Kang}}, \ and\
  \bibinfo {author} {\bibfnamefont {W.}~\bibnamefont {Cai}},\ }\href@noop {}
  {\bibfield  {journal} {\bibinfo  {journal} {Proceedings of the National
  Academy of Sciences}\ }\textbf {\bibinfo {volume} {108}},\ \bibinfo {pages}
  {5174} (\bibinfo {year} {2011})}\BibitemShut {NoStop}%
\bibitem [{\citenamefont {Kim}\ and\ \citenamefont
  {Tadmor}(2014)}]{kim2014entropically}%
  \BibitemOpen
  \bibfield  {author} {\bibinfo {author} {\bibfnamefont {W.~K.}\ \bibnamefont
  {Kim}}\ and\ \bibinfo {author} {\bibfnamefont {E.~B.}\ \bibnamefont
  {Tadmor}},\ }\href@noop {} {\bibfield  {journal} {\bibinfo  {journal}
  {Physical review letters}\ }\textbf {\bibinfo {volume} {112}},\ \bibinfo
  {pages} {105501} (\bibinfo {year} {2014})}\BibitemShut {NoStop}%
\bibitem [{\citenamefont {Grabowski}\ \emph {et~al.}(2009)\citenamefont
  {Grabowski}, \citenamefont {Ismer}, \citenamefont {Hickel},\ and\
  \citenamefont {Neugebauer}}]{grabowski2009ab}%
  \BibitemOpen
  \bibfield  {author} {\bibinfo {author} {\bibfnamefont {B.}~\bibnamefont
  {Grabowski}}, \bibinfo {author} {\bibfnamefont {L.}~\bibnamefont {Ismer}},
  \bibinfo {author} {\bibfnamefont {T.}~\bibnamefont {Hickel}}, \ and\ \bibinfo
  {author} {\bibfnamefont {J.}~\bibnamefont {Neugebauer}},\ }\href@noop {}
  {\bibfield  {journal} {\bibinfo  {journal} {Physical Review B}\ }\textbf
  {\bibinfo {volume} {79}},\ \bibinfo {pages} {134106} (\bibinfo {year}
  {2009})}\BibitemShut {NoStop}%
\bibitem [{\citenamefont {Glensk}\ \emph {et~al.}(2014)\citenamefont {Glensk},
  \citenamefont {Grabowski}, \citenamefont {Hickel},\ and\ \citenamefont
  {Neugebauer}}]{glensk2014breakdown}%
  \BibitemOpen
  \bibfield  {author} {\bibinfo {author} {\bibfnamefont {A.}~\bibnamefont
  {Glensk}}, \bibinfo {author} {\bibfnamefont {B.}~\bibnamefont {Grabowski}},
  \bibinfo {author} {\bibfnamefont {T.}~\bibnamefont {Hickel}}, \ and\ \bibinfo
  {author} {\bibfnamefont {J.}~\bibnamefont {Neugebauer}},\ }\href@noop {}
  {\bibfield  {journal} {\bibinfo  {journal} {Physical Review X}\ }\textbf
  {\bibinfo {volume} {4}},\ \bibinfo {pages} {011018} (\bibinfo {year}
  {2014})}\BibitemShut {NoStop}%
\bibitem [{\citenamefont {Allen}\ and\ \citenamefont
  {Tildesley}(2012)}]{allen2012computer}%
  \BibitemOpen
  \bibfield  {author} {\bibinfo {author} {\bibfnamefont {M.~P.}\ \bibnamefont
  {Allen}}\ and\ \bibinfo {author} {\bibfnamefont {D.~J.}\ \bibnamefont
  {Tildesley}},\ }\href@noop {} {\emph {\bibinfo {title} {Computer simulation
  in chemical physics}}},\ Vol.\ \bibinfo {volume} {397}\ (\bibinfo
  {publisher} {Springer Science \& Business Media},\ \bibinfo {year}
  {2012})\BibitemShut {NoStop}%
\bibitem [{\citenamefont {Polson}\ \emph {et~al.}(2000)\citenamefont {Polson},
  \citenamefont {Trizac}, \citenamefont {Pronk},\ and\ \citenamefont
  {Frenkel}}]{polson2000finite}%
  \BibitemOpen
  \bibfield  {author} {\bibinfo {author} {\bibfnamefont {J.~M.}\ \bibnamefont
  {Polson}}, \bibinfo {author} {\bibfnamefont {E.}~\bibnamefont {Trizac}},
  \bibinfo {author} {\bibfnamefont {S.}~\bibnamefont {Pronk}}, \ and\ \bibinfo
  {author} {\bibfnamefont {D.}~\bibnamefont {Frenkel}},\ }\href@noop {}
  {\bibfield  {journal} {\bibinfo  {journal} {The Journal of Chemical Physics}\
  }\textbf {\bibinfo {volume} {112}},\ \bibinfo {pages} {5339} (\bibinfo {year}
  {2000})}\BibitemShut {NoStop}%
\bibitem [{Note1()}]{Note1}%
  \BibitemOpen
  \bibinfo {note} {The zero-frequency translational modes are excluded due to
  the constraint on CM}\BibitemShut {NoStop}%
\bibitem [{\citenamefont {Noya}\ \emph {et~al.}(2008)\citenamefont {Noya},
  \citenamefont {Conde},\ and\ \citenamefont {Vega}}]{noya2008computing}%
  \BibitemOpen
  \bibfield  {author} {\bibinfo {author} {\bibfnamefont {E.~G.}\ \bibnamefont
  {Noya}}, \bibinfo {author} {\bibfnamefont {M.}~\bibnamefont {Conde}}, \ and\
  \bibinfo {author} {\bibfnamefont {C.}~\bibnamefont {Vega}},\ }\href@noop {}
  {\bibfield  {journal} {\bibinfo  {journal} {The Journal of chemical physics}\
  }\textbf {\bibinfo {volume} {129}},\ \bibinfo {pages} {104704} (\bibinfo
  {year} {2008})}\BibitemShut {NoStop}%
\bibitem [{\citenamefont {Ayala}\ and\ \citenamefont
  {Schlegel}(1998)}]{ayala1998identification}%
  \BibitemOpen
  \bibfield  {author} {\bibinfo {author} {\bibfnamefont {P.~Y.}\ \bibnamefont
  {Ayala}}\ and\ \bibinfo {author} {\bibfnamefont {H.~B.}\ \bibnamefont
  {Schlegel}},\ }\href@noop {} {\bibfield  {journal} {\bibinfo  {journal} {The
  Journal of chemical physics}\ }\textbf {\bibinfo {volume} {108}},\ \bibinfo
  {pages} {2314} (\bibinfo {year} {1998})}\BibitemShut {NoStop}%
\bibitem [{\citenamefont {Ceriotti}\ \emph {et~al.}(2011)\citenamefont
  {Ceriotti}, \citenamefont {Brain}, \citenamefont {Riordan},\ and\
  \citenamefont {Manolopoulos}}]{ceriotti2011inefficiency}%
  \BibitemOpen
  \bibfield  {author} {\bibinfo {author} {\bibfnamefont {M.}~\bibnamefont
  {Ceriotti}}, \bibinfo {author} {\bibfnamefont {G.~A.}\ \bibnamefont {Brain}},
  \bibinfo {author} {\bibfnamefont {O.}~\bibnamefont {Riordan}}, \ and\
  \bibinfo {author} {\bibfnamefont {D.~E.}\ \bibnamefont {Manolopoulos}},\ }in\
  \href@noop {} {\emph {\bibinfo {booktitle} {Proc. R. Soc. A}}}\ (\bibinfo
  {organization} {The Royal Society},\ \bibinfo {year} {2011})\ p.\ \bibinfo
  {pages} {rspa20110413}\BibitemShut {NoStop}%
\bibitem [{\citenamefont {Ceriotti}\ and\ \citenamefont
  {Markland}(2013)}]{ceri-mark13jcp}%
  \BibitemOpen
  \bibfield  {author} {\bibinfo {author} {\bibfnamefont {M.}~\bibnamefont
  {Ceriotti}}\ and\ \bibinfo {author} {\bibfnamefont {T.~E.}\ \bibnamefont
  {Markland}},\ }\href@noop {} {\bibfield  {journal} {\bibinfo  {journal} {J.
  Chem. Phys.}\ }\textbf {\bibinfo {volume} {138}},\ \bibinfo {pages} {014112}
  (\bibinfo {year} {2013})}\BibitemShut {NoStop}%
\bibitem [{\citenamefont {Earl}\ and\ \citenamefont
  {Deem}(2005)}]{earl2005parallel}%
  \BibitemOpen
  \bibfield  {author} {\bibinfo {author} {\bibfnamefont {D.~J.}\ \bibnamefont
  {Earl}}\ and\ \bibinfo {author} {\bibfnamefont {M.~W.}\ \bibnamefont
  {Deem}},\ }\href@noop {} {\bibfield  {journal} {\bibinfo  {journal} {Physical
  Chemistry Chemical Physics}\ }\textbf {\bibinfo {volume} {7}},\ \bibinfo
  {pages} {3910} (\bibinfo {year} {2005})}\BibitemShut {NoStop}%
\bibitem [{\citenamefont {Freitas}\ \emph {et~al.}(2016)\citenamefont
  {Freitas}, \citenamefont {Asta},\ and\ \citenamefont
  {de~Koning}}]{freitas2016nonequilibrium}%
  \BibitemOpen
  \bibfield  {author} {\bibinfo {author} {\bibfnamefont {R.}~\bibnamefont
  {Freitas}}, \bibinfo {author} {\bibfnamefont {M.}~\bibnamefont {Asta}}, \
  and\ \bibinfo {author} {\bibfnamefont {M.}~\bibnamefont {de~Koning}},\
  }\href@noop {} {\bibfield  {journal} {\bibinfo  {journal} {Computational
  Materials Science}\ }\textbf {\bibinfo {volume} {112}},\ \bibinfo {pages}
  {333} (\bibinfo {year} {2016})}\BibitemShut {NoStop}%
\bibitem [{\citenamefont {Xie}\ \emph {et~al.}(1999)\citenamefont {Xie},
  \citenamefont {de~Gironcoli}, \citenamefont {Baroni},\ and\ \citenamefont
  {Scheffler}}]{xie1999first}%
  \BibitemOpen
  \bibfield  {author} {\bibinfo {author} {\bibfnamefont {J.}~\bibnamefont
  {Xie}}, \bibinfo {author} {\bibfnamefont {S.}~\bibnamefont {de~Gironcoli}},
  \bibinfo {author} {\bibfnamefont {S.}~\bibnamefont {Baroni}}, \ and\ \bibinfo
  {author} {\bibfnamefont {M.}~\bibnamefont {Scheffler}},\ }\href@noop {}
  {\bibfield  {journal} {\bibinfo  {journal} {Physical Review B}\ }\textbf
  {\bibinfo {volume} {59}},\ \bibinfo {pages} {965} (\bibinfo {year}
  {1999})}\BibitemShut {NoStop}%
\bibitem [{\citenamefont {Ram{\'\i}rez}\ \emph {et~al.}(2012)\citenamefont
  {Ram{\'\i}rez}, \citenamefont {Neuerburg}, \citenamefont
  {Fern{\'a}ndez-Serra},\ and\ \citenamefont {Herrero}}]{ramirez2012quasi}%
  \BibitemOpen
  \bibfield  {author} {\bibinfo {author} {\bibfnamefont {R.}~\bibnamefont
  {Ram{\'\i}rez}}, \bibinfo {author} {\bibfnamefont {N.}~\bibnamefont
  {Neuerburg}}, \bibinfo {author} {\bibfnamefont {M.-V.}\ \bibnamefont
  {Fern{\'a}ndez-Serra}}, \ and\ \bibinfo {author} {\bibfnamefont
  {C.}~\bibnamefont {Herrero}},\ }\href@noop {} {\bibfield  {journal} {\bibinfo
   {journal} {The Journal of chemical physics}\ }\textbf {\bibinfo {volume}
  {137}},\ \bibinfo {pages} {044502} (\bibinfo {year} {2012})}\BibitemShut
  {NoStop}%
\bibitem [{\citenamefont {Parrinello}\ and\ \citenamefont
  {Rahman}(1981)}]{parrinello1981polymorphic}%
  \BibitemOpen
  \bibfield  {author} {\bibinfo {author} {\bibfnamefont {M.}~\bibnamefont
  {Parrinello}}\ and\ \bibinfo {author} {\bibfnamefont {A.}~\bibnamefont
  {Rahman}},\ }\href@noop {} {\bibfield  {journal} {\bibinfo  {journal}
  {Journal of Applied physics}\ }\textbf {\bibinfo {volume} {52}},\ \bibinfo
  {pages} {7182} (\bibinfo {year} {1981})}\BibitemShut {NoStop}%
\bibitem [{\citenamefont {Hoover}(1968)}]{hoover1968entropy}%
  \BibitemOpen
  \bibfield  {author} {\bibinfo {author} {\bibfnamefont {W.~G.}\ \bibnamefont
  {Hoover}},\ }\href@noop {} {\bibfield  {journal} {\bibinfo  {journal} {The
  Journal of Chemical Physics}\ }\textbf {\bibinfo {volume} {49}},\ \bibinfo
  {pages} {1981} (\bibinfo {year} {1968})}\BibitemShut {NoStop}%
\bibitem [{\citenamefont {Giannozzi}\ \emph {et~al.}(1991)\citenamefont
  {Giannozzi}, \citenamefont {De~Gironcoli}, \citenamefont {Pavone},\ and\
  \citenamefont {Baroni}}]{giannozzi1991ab}%
  \BibitemOpen
  \bibfield  {author} {\bibinfo {author} {\bibfnamefont {P.}~\bibnamefont
  {Giannozzi}}, \bibinfo {author} {\bibfnamefont {S.}~\bibnamefont
  {De~Gironcoli}}, \bibinfo {author} {\bibfnamefont {P.}~\bibnamefont
  {Pavone}}, \ and\ \bibinfo {author} {\bibfnamefont {S.}~\bibnamefont
  {Baroni}},\ }\href@noop {} {\bibfield  {journal} {\bibinfo  {journal}
  {Physical Review B}\ }\textbf {\bibinfo {volume} {43}},\ \bibinfo {pages}
  {7231} (\bibinfo {year} {1991})}\BibitemShut {NoStop}%
\bibitem [{\citenamefont {Pogatscher}\ \emph {et~al.}(2014)\citenamefont
  {Pogatscher}, \citenamefont {Antrekowitsch}, \citenamefont {Werinos},
  \citenamefont {Moszner}, \citenamefont {Gerstl}, \citenamefont {Francis},
  \citenamefont {Curtin}, \citenamefont {L{\"{o}}ffler},\ and\ \citenamefont
  {Uggowitzer}}]{poga+14prl}%
  \BibitemOpen
  \bibfield  {author} {\bibinfo {author} {\bibfnamefont {S.}~\bibnamefont
  {Pogatscher}}, \bibinfo {author} {\bibfnamefont {H.}~\bibnamefont
  {Antrekowitsch}}, \bibinfo {author} {\bibfnamefont {M.}~\bibnamefont
  {Werinos}}, \bibinfo {author} {\bibfnamefont {F.}~\bibnamefont {Moszner}},
  \bibinfo {author} {\bibfnamefont {S.~S.~A.}\ \bibnamefont {Gerstl}}, \bibinfo
  {author} {\bibfnamefont {M.~F.}\ \bibnamefont {Francis}}, \bibinfo {author}
  {\bibfnamefont {W.~A.}\ \bibnamefont {Curtin}}, \bibinfo {author}
  {\bibfnamefont {J.~F.}\ \bibnamefont {L{\"{o}}ffler}}, \ and\ \bibinfo
  {author} {\bibfnamefont {P.~J.}\ \bibnamefont {Uggowitzer}},\ }\href@noop {}
  {\bibfield  {journal} {\bibinfo  {journal} {Phys. Rev. Lett.}\ }\textbf
  {\bibinfo {volume} {112}},\ \bibinfo {pages} {225701} (\bibinfo {year}
  {2014})}\BibitemShut {NoStop}%
\bibitem [{\citenamefont {Mendelev}\ \emph {et~al.}(2003)\citenamefont
  {Mendelev}, \citenamefont {Han}, \citenamefont {Srolovitz}, \citenamefont
  {Ackland}, \citenamefont {Sun},\ and\ \citenamefont
  {Asta}}]{mendelev2003development}%
  \BibitemOpen
  \bibfield  {author} {\bibinfo {author} {\bibfnamefont {M.}~\bibnamefont
  {Mendelev}}, \bibinfo {author} {\bibfnamefont {S.}~\bibnamefont {Han}},
  \bibinfo {author} {\bibfnamefont {D.}~\bibnamefont {Srolovitz}}, \bibinfo
  {author} {\bibfnamefont {G.}~\bibnamefont {Ackland}}, \bibinfo {author}
  {\bibfnamefont {D.}~\bibnamefont {Sun}}, \ and\ \bibinfo {author}
  {\bibfnamefont {M.}~\bibnamefont {Asta}},\ }\href@noop {} {\bibfield
  {journal} {\bibinfo  {journal} {Philosophical magazine}\ }\textbf {\bibinfo
  {volume} {83}},\ \bibinfo {pages} {3977} (\bibinfo {year}
  {2003})}\BibitemShut {NoStop}%
\bibitem [{\citenamefont {Bonny}\ \emph {et~al.}(2009)\citenamefont {Bonny},
  \citenamefont {Pasianot}, \citenamefont {Castin},\ and\ \citenamefont
  {Malerba}}]{bonny2009ternary}%
  \BibitemOpen
  \bibfield  {author} {\bibinfo {author} {\bibfnamefont {G.}~\bibnamefont
  {Bonny}}, \bibinfo {author} {\bibfnamefont {R.~C.}\ \bibnamefont {Pasianot}},
  \bibinfo {author} {\bibfnamefont {N.}~\bibnamefont {Castin}}, \ and\ \bibinfo
  {author} {\bibfnamefont {L.}~\bibnamefont {Malerba}},\ }\href@noop {}
  {\bibfield  {journal} {\bibinfo  {journal} {Philosophical magazine}\ }\textbf
  {\bibinfo {volume} {89}},\ \bibinfo {pages} {3531} (\bibinfo {year}
  {2009})}\BibitemShut {NoStop}%
\bibitem [{\citenamefont {Engin}\ \emph {et~al.}(2008)\citenamefont {Engin},
  \citenamefont {Sandoval},\ and\ \citenamefont
  {Urbassek}}]{engin2008characterization}%
  \BibitemOpen
  \bibfield  {author} {\bibinfo {author} {\bibfnamefont {C.}~\bibnamefont
  {Engin}}, \bibinfo {author} {\bibfnamefont {L.}~\bibnamefont {Sandoval}}, \
  and\ \bibinfo {author} {\bibfnamefont {H.~M.}\ \bibnamefont {Urbassek}},\
  }\href@noop {} {\bibfield  {journal} {\bibinfo  {journal} {Modelling and
  Simulation in Materials Science and Engineering}\ }\textbf {\bibinfo {volume}
  {16}},\ \bibinfo {pages} {035005} (\bibinfo {year} {2008})}\BibitemShut
  {NoStop}%
\bibitem [{\citenamefont {Ceriotti}\ \emph {et~al.}(2014)\citenamefont
  {Ceriotti}, \citenamefont {More},\ and\ \citenamefont
  {Manolopoulos}}]{ceriotti2014pi}%
  \BibitemOpen
  \bibfield  {author} {\bibinfo {author} {\bibfnamefont {M.}~\bibnamefont
  {Ceriotti}}, \bibinfo {author} {\bibfnamefont {J.}~\bibnamefont {More}}, \
  and\ \bibinfo {author} {\bibfnamefont {D.~E.}\ \bibnamefont {Manolopoulos}},\
  }\href@noop {} {\bibfield  {journal} {\bibinfo  {journal} {Computer Physics
  Communications}\ }\textbf {\bibinfo {volume} {185}},\ \bibinfo {pages} {1019}
  (\bibinfo {year} {2014})}\BibitemShut {NoStop}%
\bibitem [{\citenamefont {Bussi}\ \emph {et~al.}(2007)\citenamefont {Bussi},
  \citenamefont {Donadio},\ and\ \citenamefont
  {Parrinello}}]{bussi2007canonical}%
  \BibitemOpen
  \bibfield  {author} {\bibinfo {author} {\bibfnamefont {G.}~\bibnamefont
  {Bussi}}, \bibinfo {author} {\bibfnamefont {D.}~\bibnamefont {Donadio}}, \
  and\ \bibinfo {author} {\bibfnamefont {M.}~\bibnamefont {Parrinello}},\
  }\href@noop {} {\bibfield  {journal} {\bibinfo  {journal} {The Journal of
  chemical physics}\ }\textbf {\bibinfo {volume} {126}},\ \bibinfo {pages}
  {014101} (\bibinfo {year} {2007})}\BibitemShut {NoStop}%
\bibitem [{\citenamefont {Plimpton}(1995)}]{plim95jcp}%
  \BibitemOpen
  \bibfield  {author} {\bibinfo {author} {\bibfnamefont {S.}~\bibnamefont
  {Plimpton}},\ }\href@noop {} {\bibfield  {journal} {\bibinfo  {journal} {J.
  Comp. Phys.}\ }\textbf {\bibinfo {volume} {117}},\ \bibinfo {pages} {1}
  (\bibinfo {year} {1995})}\BibitemShut {NoStop}%
\bibitem [{SI()}]{SI}%
  \BibitemOpen
  \href {to be inserted} {\enquote {\bibinfo {title} {Supplemental material},}\
  }\BibitemShut {NoStop}%
\bibitem [{\citenamefont {Allain}\ \emph {et~al.}(2004)\citenamefont {Allain},
  \citenamefont {Chateau}, \citenamefont {Bouaziz}, \citenamefont {Migot},\
  and\ \citenamefont {Guelton}}]{allain2004correlations}%
  \BibitemOpen
  \bibfield  {author} {\bibinfo {author} {\bibfnamefont {S.}~\bibnamefont
  {Allain}}, \bibinfo {author} {\bibfnamefont {J.-P.}\ \bibnamefont {Chateau}},
  \bibinfo {author} {\bibfnamefont {O.}~\bibnamefont {Bouaziz}}, \bibinfo
  {author} {\bibfnamefont {S.}~\bibnamefont {Migot}}, \ and\ \bibinfo {author}
  {\bibfnamefont {N.}~\bibnamefont {Guelton}},\ }\href@noop {} {\bibfield
  {journal} {\bibinfo  {journal} {Materials Science and Engineering: A}\
  }\textbf {\bibinfo {volume} {387}},\ \bibinfo {pages} {158} (\bibinfo {year}
  {2004})}\BibitemShut {NoStop}%
\bibitem [{\citenamefont {Van~Swygenhoven}\ \emph {et~al.}(2004)\citenamefont
  {Van~Swygenhoven}, \citenamefont {Derlet},\ and\ \citenamefont
  {Fr{\o}seth}}]{van2004stacking}%
  \BibitemOpen
  \bibfield  {author} {\bibinfo {author} {\bibfnamefont {H.}~\bibnamefont
  {Van~Swygenhoven}}, \bibinfo {author} {\bibfnamefont {P.}~\bibnamefont
  {Derlet}}, \ and\ \bibinfo {author} {\bibfnamefont {A.}~\bibnamefont
  {Fr{\o}seth}},\ }\href@noop {} {\bibfield  {journal} {\bibinfo  {journal}
  {Nature materials}\ }\textbf {\bibinfo {volume} {3}},\ \bibinfo {pages} {399}
  (\bibinfo {year} {2004})}\BibitemShut {NoStop}%
\bibitem [{\citenamefont {Yamakov}\ \emph {et~al.}(2004)\citenamefont
  {Yamakov}, \citenamefont {Wolf}, \citenamefont {Phillpot}, \citenamefont
  {Mukherjee},\ and\ \citenamefont {Gleiter}}]{yamakov2004deformation}%
  \BibitemOpen
  \bibfield  {author} {\bibinfo {author} {\bibfnamefont {V.}~\bibnamefont
  {Yamakov}}, \bibinfo {author} {\bibfnamefont {D.}~\bibnamefont {Wolf}},
  \bibinfo {author} {\bibfnamefont {S.}~\bibnamefont {Phillpot}}, \bibinfo
  {author} {\bibfnamefont {A.}~\bibnamefont {Mukherjee}}, \ and\ \bibinfo
  {author} {\bibfnamefont {H.}~\bibnamefont {Gleiter}},\ }\href@noop {}
  {\bibfield  {journal} {\bibinfo  {journal} {Nature materials}\ }\textbf
  {\bibinfo {volume} {3}},\ \bibinfo {pages} {43} (\bibinfo {year}
  {2004})}\BibitemShut {NoStop}%
\bibitem [{\citenamefont {Voter}\ and\ \citenamefont
  {Chen}(1986)}]{voter1986accurate}%
  \BibitemOpen
  \bibfield  {author} {\bibinfo {author} {\bibfnamefont {A.~F.}\ \bibnamefont
  {Voter}}\ and\ \bibinfo {author} {\bibfnamefont {S.~P.}\ \bibnamefont
  {Chen}},\ }\href@noop {} {\bibfield  {journal} {\bibinfo  {journal} {MRS
  Online Proceedings Library Archive}\ }\textbf {\bibinfo {volume} {82}}
  (\bibinfo {year} {1986})}\BibitemShut {NoStop}%
\bibitem [{\citenamefont {Hooton}(1955)}]{hooton1955li}%
  \BibitemOpen
  \bibfield  {author} {\bibinfo {author} {\bibfnamefont {D.}~\bibnamefont
  {Hooton}},\ }\href@noop {} {\bibfield  {journal} {\bibinfo  {journal} {The
  London, Edinburgh, and Dublin Philosophical Magazine and Journal of Science}\
  }\textbf {\bibinfo {volume} {46}},\ \bibinfo {pages} {422} (\bibinfo {year}
  {1955})}\BibitemShut {NoStop}%
\bibitem [{\citenamefont {Brown}\ \emph {et~al.}(2013)\citenamefont {Brown},
  \citenamefont {Georgescu},\ and\ \citenamefont {Mandelshtam}}]{brow+13jcp}%
  \BibitemOpen
  \bibfield  {author} {\bibinfo {author} {\bibfnamefont {S.~E.}\ \bibnamefont
  {Brown}}, \bibinfo {author} {\bibfnamefont {I.}~\bibnamefont {Georgescu}}, \
  and\ \bibinfo {author} {\bibfnamefont {V.~a.}\ \bibnamefont {Mandelshtam}},\
  }\href@noop {} {\bibfield  {journal} {\bibinfo  {journal} {J. Chem. Phys.}\
  }\textbf {\bibinfo {volume} {138}},\ \bibinfo {pages} {044317} (\bibinfo
  {year} {2013})}\BibitemShut {NoStop}%
\bibitem [{\citenamefont {Errea}\ \emph {et~al.}(2014)\citenamefont {Errea},
  \citenamefont {Calandra},\ and\ \citenamefont {Mauri}}]{errea2014anharmonic}%
  \BibitemOpen
  \bibfield  {author} {\bibinfo {author} {\bibfnamefont {I.}~\bibnamefont
  {Errea}}, \bibinfo {author} {\bibfnamefont {M.}~\bibnamefont {Calandra}}, \
  and\ \bibinfo {author} {\bibfnamefont {F.}~\bibnamefont {Mauri}},\
  }\href@noop {} {\bibfield  {journal} {\bibinfo  {journal} {Physical Review
  B}\ }\textbf {\bibinfo {volume} {89}},\ \bibinfo {pages} {064302} (\bibinfo
  {year} {2014})}\BibitemShut {NoStop}%
\bibitem [{\citenamefont {Campan{\'a}}\ and\ \citenamefont
  {M{\"u}ser}(2006)}]{campana2006practical}%
  \BibitemOpen
  \bibfield  {author} {\bibinfo {author} {\bibfnamefont {C.}~\bibnamefont
  {Campan{\'a}}}\ and\ \bibinfo {author} {\bibfnamefont {M.~H.}\ \bibnamefont
  {M{\"u}ser}},\ }\href@noop {} {\bibfield  {journal} {\bibinfo  {journal}
  {Physical Review B}\ }\textbf {\bibinfo {volume} {74}},\ \bibinfo {pages}
  {075420} (\bibinfo {year} {2006})}\BibitemShut {NoStop}%
\bibitem [{\citenamefont {Rossi}\ \emph {et~al.}(2015)\citenamefont {Rossi},
  \citenamefont {Fang},\ and\ \citenamefont {Michaelides}}]{ross+15jpcl}%
  \BibitemOpen
  \bibfield  {author} {\bibinfo {author} {\bibfnamefont {M.}~\bibnamefont
  {Rossi}}, \bibinfo {author} {\bibfnamefont {W.}~\bibnamefont {Fang}}, \ and\
  \bibinfo {author} {\bibfnamefont {A.}~\bibnamefont {Michaelides}},\
  }\href@noop {} {\bibfield  {journal} {\bibinfo  {journal} {J. Phys. Chem.
  Letters}\ }\textbf {\bibinfo {volume} {6}},\ \bibinfo {pages} {4233}
  (\bibinfo {year} {2015})}\BibitemShut {NoStop}%
\end{thebibliography}
\end{document}